\documentclass[10pt]{iopart}
\usepackage{iopams,graphicx}
\usepackage[jphysicsB,abbr]{harvard}
\usepackage{psfrag}
\begin{document}
\title[Superfluidity and binary-correlations]{Superfluidity and binary-correlations within clusters of fermions}

\author{J.N. Milstein and K. Burnett} \address{Clarendon Laboratory,
Department of Physics, University of Oxford, Parks Road, Oxford OX1
3PU, United Kingdom}

\begin{abstract}
We propose a method for simulating the behaviour of small clusters of particles that explicitly accounts for all mean-field and binary-correlation effects.  Our approach leads to a set of variational equations that can be used to study both the dynamics and thermodynamics of these clusters.  As an illustration of this method, we explore the BCS-BEC crossover in the simple model of four fermions, interacting with finite-range potentials, in a harmonic potential.  We find, in the crossover regime, that the particles prefer to occupy two distinct pair states as opposed to the one assumed by BCS theory.
\end{abstract}

%\submitto{\JPB}
\maketitle

\section{Introduction}
Feshbach resonances are now widely used to control the interactions between atoms in ultra-cold atomic gases.  At temperatures low enough for superfluidity to occur, these resonances allow experiments to realise the crossover between a superfluid of Cooper pairs (BCS) and a Bose-Einstein condensate (BEC) of molecules    \cite{OHara:2002ca,Chin:2004qs,Kinast:2005rb,Bourdel:2003ha,Regal:2005iy,Zwierlein:2005pn}.  Despite breathtaking advances in experiments, the task of formulating a theory justified in connecting the two extremes remains.  The difficulty stems from the increasing importance of correlations within the system and requires a theory which accounts for many-body effects at a level well beyond mean-field theory.  

There have been a variety of proposals for treating these correlation effects mostly based on the methods of diagrammatic many-body perturbation theory and relying upon the inclusion of various subsets of interaction diagrams.  Unfortunately, due to the lack of a clear expansion parameter, it is difficult to justify these approaches.  Moreover, at $T=0$, simple BCS mean-field theory smoothly interpolates throughout the BCS-BEC crossover despite it's neglect of correlations.  This behaviour was first noted by Leggett \cite{Leggett:1980cp}, but the utility of BCS theory is apparent when compared to the results of recent experiments \cite{Regal:2005iy} and Monte-Carlo calculations \cite{Astrakharchik:oc} of the crossover.  BCS theory continues to work surprisingly well within regimes where a mean-field theory is hardly justified (we will return to this point later).  At finite-temperatures it is true that conventional BCS theory diverges in the crossover, but a simple Gaussian expansion about the mean-field theory fixes this problem \cite{Nozieres:1995}.  The result is we have a theory that smoothly interpolates throughout the crossover region despite a the neglect of quantum correlations.  This surprising robustness of BCS theory has motivated us to reconsider the mean-field approach and to look for a method to extend it in a physically transparent way.

In order to understand the difficulties that correlations introduce let us recollect why mean-field theory is such an attractive approach to a many-body problem.  The allure of mean-field theory is that it lets us squeeze all the complications of the many-body problem into a single interaction term leaving us with an effective single-particle problem.  For a dilute gas, in order to include the effects of interactions among the particles, the only microscopic input needed comes from an understanding of the binary scattering problem.  Unfortunately, as higher-order correlations form throughout the system, i.e., beyond binary, mean-field theory breaks down,  and a knowledge of two-body interactions is no longer sufficient.  Rather, for a complete theory, we must account for an increasing number of N-body correlations.  A recent example that clearly illustrates this need is the four-body calculation that was necessary to find the correct molecule-molecule scattering within a dilute gas of fermions far on the BEC side of the resonance \cite{Petrov:2004} .  

Since interactions increase the number of N-body correlations that govern the microscopic physics,  a starting point for understanding the strongly-correlated regime is to better understand the physics of small clusters of atoms.  In this paper we will study a cluster of four particles, interacting with finite-range potentials, at all interaction strengths across a resonance.  This forms the simplest model one could imagine of the BCS-BEC crossover.  On the BEC side we will have  two interacting bound molecules, while on the BCS side we have Pauli blocking of the lowest energy level resulting in an analogous ``Cooper pairing."  We will attack this problem with an approach that, although approximate in its solution of the four-body problem, can be generalized to larger clusters of particles without much difficulty.  In section \ref{pairing} we discuss the basis of our approach to studying clusters, with the aid of a generalized pair wavefunction, and contrast this wavefunction with the more restrictive BCS wavefunction.  We further motivate this generalized pair wavefunction by showing how it describes both the BCS and BEC limits.  We present a variational method for working with such a wavefunction in section \ref{variational} and discuss our choice of trial pair-orbitals.  Finally, in section \ref{simulation}, we discuss our numerical results for four interacting fermions, in a harmonic trap at $T=0$, throughout the crossover regime and compare this to the BCS results.  We find that the atoms tend to form pairs of varying size within the crossover as compared to the single width of the BCS pairs.  We then discuss the relevance of this result.

\section{Binary correlations and pairing wavefunctions}
\label{pairing}
\subsection{Hartree-Fock theory and binary-correlations}
In developing a controlled method for treating correlations let us first review Hartree-Fock (HF) mean-field theory.  In the spatial representation of HF theory, the HF wavefunction is an anti-symmetric function of $N$ single particle orbitals:
\begin{equation}
\Psi_{HF}=A\{\phi_1(1)\phi_2(2)\dots\phi_N(N) \},\label{hfwf}
\end{equation}
where the operator $A$ enforces the anti-symmetry of the overall wavefunction.
A wavefunction formed from single-particle orbitals, as in equation (\ref{hfwf}), when inserted in the Schr\"odinger equation, can only account for single-particle effects, i.e., the mean-field potential felt by each particle. Correlations, on the other hand, are multiple-particle effects and may be defined as all many-body effects neglected by the HF approach. We generalise equation~(\ref{hfwf}) to account for the next order,  or all binary correlations, with an anti-symmetric function of $N/2=M$ two-particle orbitals or pair-orbitals:
\begin{equation}
\Psi_P=A\{\phi_1(1,1')\phi_2(2,2')\dots\phi_M(M,M') \}.\label{pairwf}
\end{equation}
This wavefunction looks very similar to the BCS wavefunction.  In its first quantised form, the BCS wavefunction, for a two spin mixture of fermions $\sigma=\{\uparrow,\downarrow\}$, is formed by pairing each fermion with a partner of opposite spin,  designating the pair by an orbital $\phi$, and writing the full many-body wavefunction for the $N$ fermions as the anti-symmetric product of $M=N/2$ pair-orbitals, i.e.,
\begin{equation}\label{bcs}
\Psi_{BCS}=A\{\phi(1,1')\phi(2,2')\dots\phi(M,M') \}.\label{bcswf}
\end{equation}
A remarkable feature of this wavefunction is that, despite the fact that the number of states available to each individual  particle is severely restricted by Fermi statistics,  a single pairwise orbital $\phi$ can become macroscopically occupied.  It is this feature of the many-particle wavefunction that is at the origin of the coherent behaviour associated with superfluidity.

We might, however, wonder if all the pairs will occupy just the one state.  Indeed, as the inter-particle interactions are increased, and more states within the Fermi surface become available, we might well expect to generate pairs in more than one orbital.  This additional freedom is already incorporated in the  pair wavefunction of equation~(\ref{pairwf}).  If the BCS wavefunction is thought of as a HF wavefunction built upon the Cooper pairs, the pair wavefunction may be thought to generate a ``multi-configurational" BCS theory.  The first question that we must now ask is how does one generate such a wavefunction?

For our discussion, we will treat the case of a two-spin
component gas of fermions.  We will take pair states that are
formed from a product of an even spatial wavefunction and an odd
singlet spin-orbital.  This assumption greatly simplifies the problem
of anti-symmetrization.  In fact, we are then able to write the pair
wavefunction in the following determinantal form:

\begin{equation}
\Psi_P=\sum_P\left|
\begin{array}{cccc}
\phi_{P(1)}(1,1')&\phi_{P(1)}(2,1')&\dots&\phi_{P(1)}(M,1')\\
\phi_{P(2)}(1,2')&\phi_{P(2)}(2,2')&\dots&\phi_{P(2)}(M,2')\\
\vdots&\vdots&\vdots&\vdots\\
\phi_{P(M)}(1,M')&\phi_{P(M)}(2,M')&\dots&\phi_{P(M)}(M,M')
\end{array}\right|\label{bigdet}.
\end{equation}
The sum in equation (\ref{bigdet}) denotes a summation over permutations of the
orbitals $1,2,\ldots M$.  For
example, in the case of 4-particles, the many-body pair wavefunction
may be written as:
\begin{eqnarray}
\fl\Psi=\phi_1(1,1')\phi_2(2,2')+\phi_1(2,2')\phi_2(1,1')-\phi_1(1,2')\phi_2(2,1')-\phi_1(2,1')\phi_2(1,2').\label{4partwave}
\end{eqnarray}

The pair wavefunction of equation~(\ref{pairwf}) was explored extensively  in the early days of superconductivity to explain the phenomena in terms of ``Schafroth condensation," which is an analogy to Bose condensation for pairs of fermions.  This wavefunction is also an excellent candidate for studying the BCS-BEC crossover since it can be shown to correctly tend to both the weakly interacting limits of the crossover {\it without restricting the particles to occupy only a single orbital} \cite{Blatt:1964} .  Unfortunately, equation~(\ref{pairwf}) is rather difficult to work with since the number of determinants needed to generate the wavefunction scales like $M!$.  This difficulty will ultimately limit the number of particles that we can study with such a trial wavefunction.    However, since we are only concerned with small clusters of Fermions, equation~(\ref{pairwf}) seems an excellent choice for a trial wavefunction.

As mentioned above, the pair wavefunction is similar in form to the Hartree-Fock (HF) wavefunction.  The HF wavefunction, however, is an anti-symmetrized product of single-particle orbitals and neglects all correlations apart from those due to the Pauli exclusion principle.  In contrast, the pair wavefunction is an anti-symmetrized product of two-particle orbitals and may be roughly thought of as a second order HF wavefunction which accounts for the additional effect of all two-particle correlations.  We now show how this trial wavefunction can treat both the BCS and BEC limits.

\subsection{BEC limit}\label{becsec}
To illustrate the utility of equation~(\ref{pairwf}), we shall examine the four-particle case, borrowing much of the following
discussion from Blatt \cite{Blatt:1964}.  To begin, we should note that the
norm of the four-particle wavefunction has three
contributions i.e.,

\begin{equation}
\langle\Psi|\Psi\rangle=\int |\Psi|^2d^6x d^6y=4(N+D-S).\label{normex}
\end{equation}
The first contribution, $N=\langle\phi_1|\phi_1\rangle\langle\phi_2|\phi_2\rangle$, arises from the product of the norms of the pair-orbitals.  The remaining two terms come from
exchange processes inherent in the wavefunction. When both a spin up and a
spin down are exchanged between pairs, the
resulting contribution to the norm is given by the term $D=|\langle\phi_1|\phi_2\rangle|^2$
which is the modulus square of the overlap between the two pairs.  If only a single up or down spin is exchanged, however, we get the following contribution which we explicitly write:
\begin{equation}\label{singlex}
 S=\int \phi^*_1(1,1')\phi^*_2(2,2')\left(\phi_1(1,2')\phi_2(2,1')+\phi_1(2,1')\phi_2(1,2')\right) d^6xd^6y.
\end{equation}

To understand the importance of these two exchange terms, let us consider the following.  Assume for the moment that the integral of equation (\ref{singlex}) gives zero.  The resulting norm of the many-body wavefunction is identical to the result we would have attained had we started with
the much simpler wavefunction
\begin{equation}\label{bosewf}
\Psi_B(1,1';2,2')\approx\phi_1(1,1')\phi_2(2,2')+\phi_1(2,2')\phi_2(1,1').
\end{equation}
Note, this wavefunction is now symmetric which means that we are
in a regime where the fermion pairs behave like bosons.  This shows that the contribution from $S$ can be considered a measure of how much the pairs deviate from being bosons.  The only place we would expect to find $S=0$ is in the BEC limit. The moment we move away from the asymptotic limit of BEC, this term begins to grow which means that a bosonic approximation to the pairing should in general be poor. Note, this also implies that two pairs do not scatter like two bosons, even when working with the BCS approximation to the wavefunction.  

\subsection{BCS limit}
The pairing wavefunction of
equation~(\ref{pairwf}), which can describe the Bose condensed limit, is also
appropriate in the limit of weak, attractive interactions where BCS theory should hold.  
This is clearly seen in the single orbital limit
(i.e., $\phi_1=\phi_2=\ldots\phi_P$), equation~(\ref{bcs}), when the pair-orbital is given by
\begin{equation}\label{bcsorbit}
\phi(x,y)=\sum_k g_k\e^{i k |x-y|}.
\end{equation}
For the non-interacting Fermi gas, the Fourier component $g_k$  represents a Fermi sea filled to Fermi wavenumber $k_f$:
\begin{equation}
g_k={\Bigg\{}
\begin{array}{cc}
1& |k|<k_f\\
0& |k|>k_f
\end{array}.
\end{equation}
The form of the pair-orbitals that arise in the superfluid state are virtually
unchanged from the non-interacting case.  The Fourier components of the orbitals only
change for those with wave numbers close to the Fermi surface.  This
subtle difference results in a long-range tail to the pair-orbitals that
correlate two fermions of opposite spin.  These pairs have such
a large average size that a great many of them may overlap which is in stark contrast to a dilute BEC of molecules.

Despite the very different nature of a BCS superfluid from a BEC, we see that
once again a wavefunction of the form of equation (\ref{pairwf}) is
appropriate.  As a final illustration of this point, we again refer to the case of  four particles.  Let us assume that, within the weakly-attractive limit, each pair-orbital asymptotes to a product of two
single particle orbitals
\begin{eqnarray}
\phi_1(x,y)\rightarrow\varphi_1(x)\varphi_{1'}(y)&,&
\phi_2(x,y)\rightarrow\varphi_2(x)\varphi_{2'}(y).
\end{eqnarray}
Here we use the notation $\varphi_\sigma$ to denote single particle
orbitals.  In this limit, equation~(\ref{4partwave}) may be written as
\begin{eqnarray}
\Psi=
\left|\begin{array}{cc}
\varphi_1(1)&\varphi_1(2)\\
\varphi_2(2)&\varphi_2(2)
\end{array}\right|
\left|\begin{array}{cc}
\varphi_{1'}(1')&\varphi_{1'}(2')\\
\varphi_{2'}(2')&\varphi_{2'}(2')
\end{array}\right|,
\end{eqnarray}
which is again the result for the non-interacting case. 
\section{Variational approach to the dynamics/thermodynamics}
\label{variational}
\subsection{Variational dynamical equations}
In the previous sections we have
motivated the use of a pair wavefunction to describe superfluidity in
both the strong and weak coupling limits.  We must now address how
such a wavefunction may be used to solve the many-body problem.  In theory, one could self-consistently solve for the pair-orbitals and use the resulting wavefunction to calculate any desired expectation value.  We will not perform such a calculation here, instead, we have chosen a simpler variational method of obtaining the pair-orbitals.  We follow this approach for two reasons.  First, a variational calculation is much simpler to implement than a full, self-consistent solution.  Second, and more importantly, the variational approach we now describe will allow us to generalise our method to finite-temperatures.  

We begin our discussion by writing down the time-dependent Schr\"odinger
equation in variational form
\begin{equation}
S=\int_{t_1}^{t_2} dt\langle Q(t)|i\hbar \frac{d}{dt}-\hat
H|Q(t)\rangle.\label{action}
\end{equation}
If we identify the kernel above as a Lagrangian, we are able to make a
rather elegant connection with classical mechanics (for an excellent review
of this subject see reference \cite{Feldmeier:2000eb}).  Let us divide the Lagrangian in two parts as follows:
\begin{eqnarray}
{\cal L}(\dot Q(t),Q(t))&=&\langle Q(t)|i\hbar
\frac{d}{dt}|Q(t)\rangle-\langle Q(t)|\hat H|Q(t)\rangle\nonumber
\\ &=& {\cal
L}_0(\dot Q(t),Q(t))-{\cal H}(Q(t)).
\end{eqnarray}
The wavefunction is then parametrized in terms of the set of variational
parameters $\{\dot Q(t),Q(t)\}=\{\dot q_1(t),\dot q_2(t)\dots \dot
q_N(t);q_1(t),q_2(t)\dots q_N(t)\}$.  Minimising 
equation~(\ref{action}) results in the Euler-Lagrange equations of motion
for the variational parameters
\begin{equation}\label{eleq}
\frac{d}{dt}\frac{\partial \cal L}{\partial \dot q_i}-\frac{\partial
\cal L}{\partial q_i}=0.
\end{equation}
It is often more convenient to have the Euler-Lagrange equations in a form which
only depends upon derivatives with respect to the parameters $q_\sigma$
and $\dot q_\sigma$ :
\begin{equation}\label{heq}
\sum_\sigma A_{\nu \sigma}\dot q_\sigma=-\frac{\partial\cal H}{\partial q_\nu},
\end{equation}
where $A_{\nu \sigma}$ is a skew symmetric matrix defined as
\begin{equation}
 A_{\nu \sigma}\equiv \frac{\partial^2{\cal L}_0}{\partial \dot
 q_\nu\partial q_\sigma}-\frac{\partial^2 {\cal L}_0}{\partial\dot q_\sigma
 \partial q_\nu}.
\end{equation}
The solution of equations (\ref{eleq}) or (\ref{heq}) completely specifies the time-evolution of the quantum mechanical many-body problem.  Besides providing a set of dynamical equations for the quantum mechanical system, these equations may be used to study the system at finite-temperatures. By working within a microcanonical ensemble, assuming that the system is sufficiently ergodic, we may replace thermal averages with time averages.  The evolution of the state $|Q(t)\rangle$ may be reconstructed from the variational parameters and then used to evaluate any desired expectation values.  In this paper, however, we limit ourselves to the zero temperature, equilibrium case so will not concern ourselves with the time evolution of the parameters.  

\subsection{Trial orbitals}
For the special case we are concerned with, that of a superfluid system, we will use the trial state $|Q\rangle=|\Psi\rangle$,
where the spatial representation of the ket $|\Psi\rangle$ is given by (\ref{4partwave}).  The pair-orbitals which construct our many-body
wavefunction are fully parametrized by the variables $q_\sigma$:
\begin{equation}
|\phi\rangle=|\phi(q_1,q_2\dots q_N)\rangle.
\end{equation}
The efficacy of this approach is dependent upon a good choice of trial pair-orbitals.  For our present study, we make a very simple choice for the pair-orbitals which is the set of unnormalised Gaussians of the form:
\begin{equation}\label{gorbitals}
\langle{\bf x},{\bf y}|\phi_\sigma(t)\rangle=\exp\left(\frac{|{\bf x}-{\bf
 y}|^2}{2 a^2_\sigma}-\frac{({\bf
 x}+{\bf y})^2}{2 A^2_\sigma}\right).
\end{equation}
The variables $a_\sigma(t)$ ($A_\sigma(t)$) parameterise the relative (c.m.) width of the pairs at positions $\mathbf{x}$ and $\mathbf{y}$. We chose Gaussian functions so the wavefunctions become exact in both the non-interacting Bose and Fermi limits.  Of course, a Gaussian decays much faster than the $\varphi_b\sim\exp(-x/\tau)/r$ form expected for a bound-state near threshold \cite{Taylor:1972pb}.  The quantitative errors resulting from this choice of wavefunction will be greatest when we approach the resonance where the exponentially decaying tail of the Gaussian is unable to account for the actual spatial decay of the wavefunction.  However,  in the the strongly-interacting region, i.e., near resonance, we believe that Gaussian orbitals are again adequate.  Although the binary scattering near threshold predicts a long tail on the wavefunction, the only length scale that should effect the many-body system is the inter-particle spacing.  Since our results connect smoothly between the unitary and weakly-interacting regimes we believe that our choice for the orbitals will be able to display qualitatively accurate features of the crossover.

\section{Four-particle BCS-BEC crossover}\label{simulation}
\subsection{Four-particle model}
We performed a numerical minimisation of the variational energy
\begin{equation}\label{varenergy}
E=\frac{\langle\Psi | \hat H | \Psi \rangle}{\langle\Psi | \Psi \rangle},
\end{equation}
where the four-particle Hamiltonian is defined as
\begin{eqnarray}
\fl\hat H=\sum_i\left(-\frac{\hbar^2}{2m}(\nabla_{{\bf x}_i}^2+\nabla_{{\bf y}_i}^2)+\frac{1}{2}m\omega^2({\bf x}_i^2+{\bf y}_i^2)\right)+\sum_{i< j}V(|{\bf x}_i-{\bf y}_j|).
\end{eqnarray}
The trapping frequency is given by $\omega$ and we only allow for interactions between spin $\uparrow$ and $\downarrow$ particles.  We made use of a Gaussian potential of the form: $V(|{\bf x}_i-{\bf y}_j|)=-U e^{-|{\bf x}_i-{\bf y}_j|^2/\sigma^2}$, where $U$ gives the depth of the potential and $\sigma$ represents the width.  A Gaussian is chosen simply because it separates for each dimension, a fact that will greatly aid in the numerics.  To tune the interaction we set the width $\sigma$ to a given value and adjust the depth $U$ of the potential about it's first bound state which generates an effective threshold scattering resonance.  The s-wave scattering length has the dispersive form $a_{sc}=-\pi^{1/2}\tilde U\sigma^3/(4(1-\tilde U \sigma^2/2))$, and we have absorbed the remaining physical constants into $\tilde U=(m/\hbar^2)U$.

The minimisation is straightforward to perform using a multi-dimensional quasi-Newton method \cite{Press:2002qv}.  The greatest difficulty posed by equation~(\ref{varenergy}), however, is in the evaluation of the matrix elements.  Each integral must be evaluated over four particles as opposed to two which effectively doubles the dimensionality.  Various exchange and spatial symmetries help to reduce the dimensionality of the problem, nonetheless, the matrix elements remain the most inefficient part of our program.  This has imposed two constraints: I.) We have to work with a rather large width to the potential so we can integrate over enough points within the range of the potential.  This, of course, degrades the universality of the potentials we have chosen. II.)  The width of the orbitals cannot get much smaller than the width of the potentials or again we will not be able to accurately evaluate the integrals.  Ultimately, this means that the distance we can move below the resonance is limited by our grid spacing.  Both of these problems can perhaps be circumvented by a more efficient evaluation of the matrix elements, e.g., through the use of a non-linear grid or an adaptive form of Gaussian-quadrature.  Our current approach should, however, be sufficient to justify our qualitative results.

We first confirmed that the orbitals of equation~(\ref{gorbitals}) result in $8\hbar\omega$ for the non-interacting ground state energy.  By dropping terms related to the exchange of a single fermion between the pair of orbitals, we could also reach the non-interacting Bose limit $6\hbar\omega$ (as explained in section \ref{becsec}).  After performing these baseline tests, we fixed the width of the interaction potential at $\sigma=0.3a_0$, where $a_0=\sqrt{\hbar/m\omega}$ is the oscillator length.  We chose this value of $\sigma$ as the minimum value for which we could accurately perform the matrix elements.

\begin{figure}[t]
\begin{center}
\psfrag{ylabel1}[b1]{Ground State Energy $(E-2\epsilon_b)/\hbar\omega$}
\psfrag{xlabel1}[t1]{$a_0/a_{sc}$}
\psfrag{ylabel2}[B1]{$|E-E_{BCS}|$}
\psfrag{xlabel2}[B1]{$a_0/a_{sc}$}
\includegraphics[width=10cm]{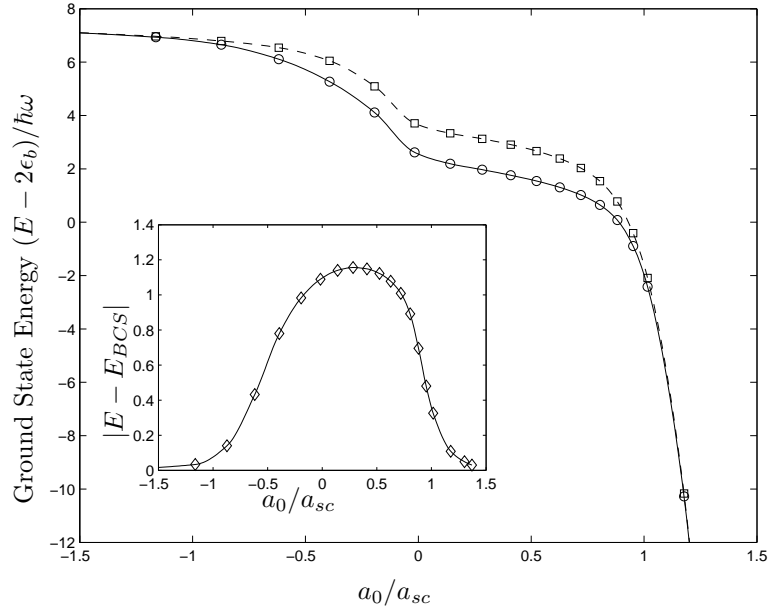}
\fl\caption{The main graph plots the total energy minus the binding energy of each pair $\epsilon_b$ as a function of the inverse scattering length.  The BCS result ({\tiny $\square$}) is plotted with that of the two-orbital, pairing model ($\circ$). The insert shows the relative difference between the two energies.  All energies are in units of the trap energy $\hbar\omega$ and lengths are in units of the oscillator length \mbox{$a_0$}. \label{fig1}}
\end{center}
\end{figure}

\subsection{Numerical results}

In the following analysis, we refer to the results obtained with the pair wavefunction of equation~(\ref{pairwf}) as the ``pairing model" while the ``BCS model" refers to the use of equation~(\ref{bcswf}).  For the four-particle problem, the crucial difference between the two models is that, while the BCS model restricts the pairs to a single orbital, the pairing model allows for the occupation of two separate orbitals.  Before discussing our numerical results, we should make two points.  First, the variational orbitals we have chosen will modify the zero of the resonance, i.e., the point at which a two-body bound state first appears.  To account for this shift we calculate the potential depth where the first bound state arises, and then simply shift the zero of the resonance accordingly.  The scattering length is then given by the theoretical value $a_{sc}$ about this point.  For the Gaussian orbitals we have chosen, of range $\sigma=0.3 a_0$, the bound state appears at a depth of $U_b\sim3.51\hbar\omega$.  Note, we work in terms of the dimensionless length $a/a_0$ as opposed to $k_f a$.  For four particles the Fermi vector $k_f$ is somewhat arbitrary whereas $a_0$ seems the more natural unit of length.  Second, our results clearly show that, as we move toward weak-repulsive interactions,  the four-particle cluster collapses.  This will prevent us from reaching the dilute BEC regime, a result which was perhaps expected given the large spatial extent of our inter-particle potentials.  One could argue that this collapse behaviour is quite physical and that the experimentally observed crossover to a dilute BEC is only quasi-stable, the four particles eventually solidify and falling from the trap.  We could perhaps simulate this quasi-stability by using very short range potentials so that the minimisation algorithm has a difficult time finding the true ground state but rather leads us to the local minimum of a quasi-stable, dilute BEC of molecules.  

 In figure~(\ref{fig1}) we compare the ground state energy of the pairing model, after subtracting off the variational bound state energy of the pairs, to the same quantity as given by the BCS model .   As the resonance is approached from above, meaning we begin on the BCS side of the crossover,  both models initially predict the same ground state energy.  The pairing model, however, branches off toward lower energies, only to converge again with the BCS model well below the resonance.  The insert of figure~(\ref{fig1}) plots the difference between these two curves to emphasise their deviation which is notably peaked just below the resonance.  It  should be noted that both curves display a similar, linear behaviour in the universal regime near the resonance.  Unfortunately, the collapse of the cluster, which drags all our curves downward as we move toward the BEC side of the resonance, prevents us from making a qualitative comparison between our model and other predictions for the equation of state within the unitarity limit.  The energy continues to shoot toward negative values below the resonance signalling a complete collapse of the cluster.
 
 \begin{figure}[t]
\begin{center}
\psfrag{ylabel1}[b1]{relative pair width $a/a_0$}
\psfrag{xlabel1}[t1]{$a_0/a_{sc}$}
\psfrag{ylabel2}[B1]{c.m. pair width $A/a_0$}
\psfrag{xlabel2}[B1]{$a_0/a_{sc}$}
\includegraphics[width=13cm,height=6cm]{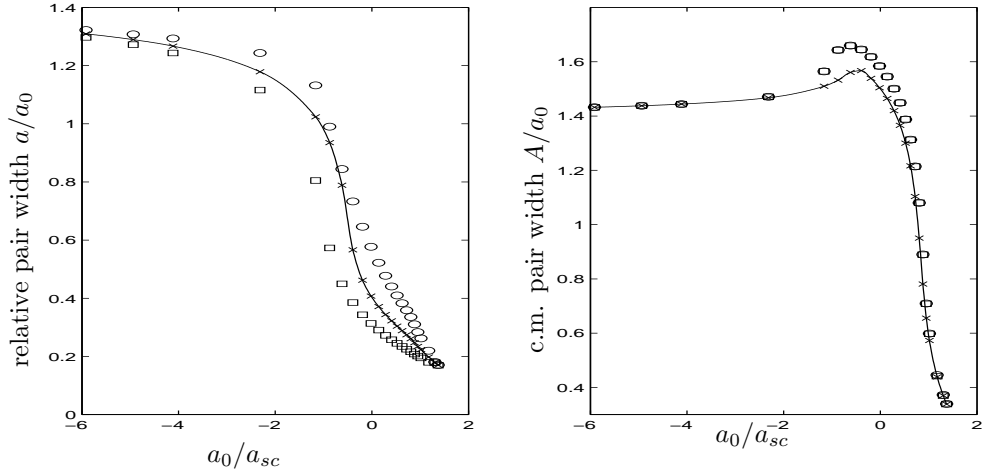}
\fl\caption{The graph on the left shows the relative width of the pairs, given by the variational parameter $a$, as a function of inverse scattering length.  The BCS result ($*$) is fitted with a line to guide the eye.  The widths of the two orbitals of the pairing wavefunction are given by the symbols {\tiny$\square$} and $\circ$.  The graph on the right shows the c.m. width of the pairs which is given by the variational parameter $A$.  The symbols are the same as in the leftmost graph (note, the circles and squares overlap).  Again, all lengths are in units of the oscillator length \mbox{$a_0$}. \label{fig2}}
\end{center}
\end{figure}

To better visualise the four-particle wavefunction, in figure~(\ref{fig2}) we plot the relative and c.m. widths as a function of inverse scattering length.  The relative width serves as a measure of the size of each pair.  If we move toward the resonance from the BCS side, one orbital is preferred by both pairs of atoms, which is rather large and stretches across the trap.  As we continue to approach the resonance, the pairs separate into two orbitals with one larger and one smaller than the BCS orbital.  Both orbitals gradually shrink as we move through the resonance only to converge well below the resonance at which point they are on the order of the size of the potential. We also see from the figures that the c.m. widths of the pairs, in the pairing model, tend to spread more within the crossover region than does the BCS orbital which is what we might expect to accompany the diminishing Fermi sea.  Both pairs, however, tend to lock onto the same c.m. width throughout the crossover.  The c.m. width quickly tends toward zero as we move below the resonance which is another sign that our cluster is collapsing.

\section{Conclusions and Outlook}
\label{conclusions}
We have developed a variational method for studying small clusters of correlated fermions that accounts for all binary-correlations among the particles.  The method is based upon a very general form of the many-particle wavefunction, similar to BCS, but not as restrictive in that it allows each pair to occupy a separate orbital. We have described in detail a variational method that can be used to study the equilibrium, dynamical and finite-temperature behaviour of these clusters using this wavefunction as an ansatz.  As an example, we have applied our method to the study of the equilibrium properties of the BCS-BEC crossover for four particles, interacting through finite-range potentials, in a harmonic trap.   We see signatures of universal behaviour for our four-particle system near the resonance, but are restricted from moving into the dilute BEC regime by the collapse of the cluster.  We have found that the system fragments from the BCS prediction of a single pair to two pairs, one larger and the other smaller than the BCS pair.  The presence of distinct pairs at $T=0$ should be contrasted with the idea of  ``preformed" or ``noncondensed" pairs that is often discussed in relation to the crossover.  These pairs are often thought to arise from finite-temperature effects or fluctuations on the order parameter.  Here, however, the distinct pairs that form within the crossover region are truly a ground state property of the system at zero temperature.

There are a few final points we should emphasise. The behaviour of the pairs, as displayed in figure~(\ref{fig2}), hints at why BCS theory interpolates throughout the BCS-BEC crossover so well since the BCS result tends to act as an average of the various pairs that form.  However, one would need to look at larger clusters to see if this pattern holds up.  It is clear, nonetheless, that by allowing the pairs to occupy different orbitals one can better model the true ground state.  Although the full pairing wavefunction of equation~(\ref{pairwf}) is much too complicated to work with for large numbers of particles, the inclusion of a restricted set of orbitals could be implemented as the variational state within quantum Monte-Carlo (MC) calculations.  It would be of interest to see, for instance, the effect of an additional pair-orbital on the fixed-node MC results when a BCS wavefunction is chosen as the variational state determining the location of the nodes.  As a final note, we feel that the results of the present work, due to the rather large width of the potentials, might be ideally suited for studying systems such as liquid ${\rm He^4}$ where the range of the potential is on the order of the inter-particle spacing.  In these systems, the collapse we see of the clusters may be unavoidable and therefore of more interest than the quasi-stable, crossover state that is the current concern in dilute gases.

\ack
J. N. Milstein is supported by a Royal Society USA Research Fellowship and  K. Burnett acknowledges funding from both the Royal Society of London and the Wolfson foundation.

\section*{References}
   \bibliography{ffff}

\end{document}